\documentstyle[epsfig]{mn}

\newif\ifAMStwofonts

\ifoldfss
  \ifCUPmtlplainloaded \else
    \NewTextAlphabet{textbfit} {cmbxti10} {}
    \NewTextAlphabet{textbfss} {cmssbx10} {}
    \NewMathAlphabet{mathbfit} {cmbxti10} {} 
    \NewMathAlphabet{mathbfss} {cmssbx10} {} 
  \fi
  \ifAMStwofonts
    \ifCUPmtlplainloaded \else
      \NewSymbolFont{upmath} {eurm10}
      \NewSymbolFont{AMSa} {msam10}
      \NewMathSymbol{\upi}     {0}{upmath}{19}
      \NewMathSymbol{\umu}     {0}{upmath}{16}
      \NewMathSymbol{\upartial}{0}{upmath}{40}
      \NewMathSymbol{\leqslant}{3}{AMSa}{36}
      \NewMathSymbol{\geqslant}{3}{AMSa}{3E}

    \fi
  \fi
\fi 

\ifnfssone
  \newmathalphabet{\mathit}
  \addtoversion{normal}{\mathit}{cmr}{m}{it}
  \addtoversion{bold}{\mathit}{cmr}{bx}{it}
  \newmathalphabet{\mathbfit} 
  \addtoversion{normal}{\mathbfit}{cmr}{bx}{it}
  \addtoversion{bold}{\mathbfit}{cmr}{bx}{it}
  \newmathalphabet{\mathbfss} 
  \addtoversion{normal}{\mathbfss}{cmss}{bx}{n}
  \addtoversion{bold}{\mathbfss}{cmss}{bx}{n}
  \ifAMStwofonts
    \ifCUPmtlplainloaded \else
      \UseAMStwoboldmath
      \makeatletter
      \new@mathgroup\upmath@group
      \define@mathgroup\mv@normal\upmath@group{eur}{m}{n}
      \define@mathgroup\mv@bold\upmath@group{eur}{b}{n}
      \edef\UPM{\hexnumber\upmath@group}
      \new@mathgroup\amsa@group
      \define@mathgroup\mv@normal\amsa@group{msa}{m}{n}
      \define@mathgroup\mv@bold\amsa@group{msa}{m}{n}
      \edef\AMSa{\hexnumber\amsa@group}
      \makeatother
      \mathchardef\upi="0\UPM19
      \mathchardef\umu="0\UPM16
      \mathchardef\upartial="0\UPM40
      \mathchardef\leqslant="3\AMSa36
      \mathchardef\geqslant="3\AMSa3E
    \fi
  \fi
\fi 

\ifnfsstwo
  \DeclareMathAlphabet{\mathbfit}{OT1}{cmr}{bx}{it}
  \SetMathAlphabet\mathbfit{bold}{OT1}{cmr}{bx}{it}
  \DeclareMathAlphabet{\mathbfss}{OT1}{cmss}{bx}{n}
  \SetMathAlphabet\mathbfss{bold}{OT1}{cmss}{bx}{n}
  \ifAMStwofonts
    \ifCUPmtlplainloaded \else
      \DeclareSymbolFont{UPM}{U}{eur}{m}{n}
      \SetSymbolFont{UPM}{bold}{U}{eur}{b}{n}
      \DeclareSymbolFont{AMSa}{U}{msa}{m}{n}
      \DeclareMathSymbol{\upi}{0}{UPM}{"19}
      \DeclareMathSymbol{\umu}{0}{UPM}{"16}
      \DeclareMathSymbol{\upartial}{0}{UPM}{"40}
      \DeclareMathSymbol{\leqslant}{3}{AMSa}{"36}
      \DeclareMathSymbol{\geqslant}{3}{AMSa}{"3E}
    \fi
  \fi
\fi 

\ifCUPmtlplainloaded \else
  \ifAMStwofonts \else 
    \def\upi{\pi}
    \def\umu{\mu}
    \def\upartial{\partial}
  \fi
\fi

\title{Magnetic Confinement, MHD Waves, and Smooth Line Profiles in AGN}
\author[M. C. Bottorff and G. J. Ferland]
       {M. C. Bottorff and Gary J. Ferland \\
        Department of Physics \& Astronomy, University of Kentucky,
        Lexington, KY 40506-0055}
\date{Accepted 2000 ??.
      Received 1999 November 10}

\pagerange{\pageref{firstpage}--\pageref{lastpage}}
\pubyear{1994}

\begin{document}

\maketitle

\label{firstpage}
\begin{abstract}

\noindent In this paper, we show that if the broad line region clouds are in
approximate energy equipartition between the magnetic field and gravity, as
hypothesized by Rees, there will be a significant effect on the shape and
smoothness of broad emission line profiles in active galactic nuclei.  Line
widths of contributing clouds or flow elements are much wider than their thermal
widths, due to the presence of non-dissipative MHD waves, and their collective contribution
produce emission line profiles broader and smoother than would be expected if a
magnetic field were not present.  As an illustration, a simple model of
isotropically emitting clouds, normally distributed in velocity, is used to show
that smoothness can be achieved for less than $\sim8\times 10^{4}$ clouds and
may even be as low as a few hundred.  We conclude that magnetic confinement has
far reaching consequences for observing and modeling active galactic nuclei.

\end{abstract}

\section{Introduction}

Quasar emission lines are important probes of the physics of active galactic
nuclei (AGN).  The primary assumption made in spectral line synthesis studies of
AGN is that the width of a component contributing to the total line profile is
thermal.  The assumption is made regardless of whether the component is a single
cloud in a discrete ensemble of clouds or a differential volume element in a
continuous flow.

Line emission originates in matter at
$\sim 10^{4}$~K and corresponds to a thermal width of $\sim 10
\rm{~km~s^{-1}}$.  Since broad line region (BLR) emission lines have much
larger widths (FWHM $\sim 10^{3}$ to $\sim10^{4}~\rm{km~s^{-1}}$) the width
of an individual component is often assumed to be negligible compared to the
total line profile.  Based on this assumption and the observed smoothness of
broad emission line profiles, cloud numbers in excess of $10^{7}$ to $10^{8}$
are inferred (Arav et al.\ 1997, Arav et al.  \ 1998, Dietrich et al.  \ 1999).

This need not be the case however.  If a magnetic field is present, the line
width of a contributing element will be broadened requiring fewer clouds
to produce the observed profile smoothness.

In nature a magnetic field is
usually associated with non-dissipative MHD waves in energy equipartition with
the magnetic field.  Thus
\begin{equation}
\frac{B^{2}}{8\pi}\approx \frac{1}{2}\rho \sigma_{\rm{B}}^{2}
\end{equation} 
where $B^{2}/8\pi$ and $1/2\rho \sigma_{\rm{B}}^{2}$ are the magnetic pressure and
MHD wave energy density respectively, $\rho$ is the mass density and $\sigma_{\rm{B}}$
is the resulting velocity width of the gas.  (Arons and Max 1974, Meyers and
Goodman~1988a, Meyers and Goodman~1988b).

Rees (1987) suggested that BLR clouds are 
magnetically confined. Assuming
\begin{equation}
\frac{B^{2}}{8\pi}\ga nkT
\end{equation}
and solving for $B$ gives
\begin{equation}
B\ga\sqrt{8\pi nkT} \sim 0.6\sqrt{n_{10}T_{4}} \rm{~G}
\end{equation}
where $n_{10}$ is the density in units of $10^{10}\rm{~cm^{-3}}$ and
$T_{4}$ is the temperature in units of $10^{4}\rm{~K}$.  From this
we see that only a few Gauss are required for confinement. Substitution
of Equation~3 into Equation~1 and solving for $\sigma_{\rm{B}}$ gives
a lower bound for the line width of magnetically confined BLR gas.  Thus
\begin{equation}
\sigma_{\rm{B}}\ga \frac{B}{\sqrt{4\pi \rho}} \approx \sqrt{\frac{2kT}{m_{A}Z}}
\approx 11 \sqrt{T_{4}} \rm{~km~s^{-1}}
\end{equation}
where $m_{A}$ is one atomic mass unit and $Z$ is the mean atomic weight of the
gas which, assuming cosmic abundances, is taken to be $Z\approx1.4$.  This is
comparable to the thermal width of hydrogen, but is roughly 3.5 times larger than
the thermal width of carbon ($\sim 3.2 \rm{~km~s^{-1}}$), and 7.5 times larger
than the thermal width of iron ($\sim 1.5 \rm{~km~s^{-1}}$) at
$T_{4}\approx1.0$.  Thus, even for the minimal confining magnetic field there will
be significant effects on line transfer.  

The broadening is considerably amplified
however if the magnetic field is in equipartition with the gravitational energy
density so that
\begin{equation}
\frac{B^{2}}{8\pi}\approx \frac{GM\rho}{R}.
\end{equation}
(Blandford and Payne 1982, Rees 1987, Emmering et al.\ 1992,
K\"onigl and Kartje 1994, Bottorff et al.\ 1997).
Although there is no fundamental reason why Equation~5 should hold, this
equipartition does occur in many environments 
(Rees (1987), Meyers and Goodman (1988a,b)).
Here $R$ is the radial distance of gas, with mass density $\rho$, from a central
black hole of mass $M$.  In this case, the line width 
will be greater than the virial width, the local velocity field will
be highly supersonic and MHD broadening may actually account for the full width
of an emission line.

\section[]{A Simple Model}

To illustrate the effect of non-dissipative MHD wave line broadening and
smoothing on an emission line profile, we consider the extreme case of BLR
emission arising from a discrete set of identical clouds.  This example has
immediate applicability in the search for discrete clouds or extended cloud
structures in BLR line profiles.  (Arav et al.\ 1997, Arav et al.  \ 1998,
Dietrich et al.  \ 1999)

\subsection[]{A Discrete Distribution of Emitters}

A simple one dimensional outflow model is constructed in which clouds 
move only along the line of sight. 
We choose a cloud bulk velocity field given by
\begin{equation}
v_{G}\approx \sqrt{\frac{2GM}{R}}.
\end{equation}
Requiring consistency with Equation~1 and Equation~5 gives
\begin{equation}
\sigma_{\rm{B}} \approx \sqrt{\frac{B^{2}}{4\pi \rho}}\approx \sqrt{\frac{2GM}{R}}
\approx v_{G}.
\end{equation}
Thus the dispersion of an emitting element equals the systemic velocity at any
given radius.  (Note:  A magnetic example in which 
$v_{G}\sim \sqrt{2GM/R}$ is the MHD wind model of Blandford and Payne (1982). In 
that paper $v_{G}$ is also given by $v_{G}\sim \sqrt{B^{2}/4\pi \rho}$. An
association of $\sigma_{\rm{B}}$ with $v_{G}$, however, is not pursued.)  

In the model being considered here identical clouds are placed randomly 
in velocity space according to a gaussian distribution given by
\begin{equation}
f(v_{G}/\sigma_{G})\propto \exp{-\frac{1}{2}\left( \frac{v_{G}}{\sigma_{G}}\right)^2}
\end{equation}
where $\sigma_{G}$ is the dispersion of the cloud
distribution in velocity space.  We loosely associate $\sigma_{G}$ with the
mass $M$ and an emission weighted radius $R_{G}$ (e.g.\ the reverberation
radius) giving $\sigma_{G} \approx \sqrt{2GM/R_{G}}$.  Thus $\sigma_{G}$ does
not include the effect of magnetic broadening which must be added separately to
each cloud.

We will predict the profile of the 1549\AA ~line of C{\sc~iv}.
The surface emissivity of 
a cloud, $\epsilon(C{~\sc iv})$, is given by an analytical fit of C{~\sc iv}
emission for an amalgam of BLR cloud densities as prescribed in Baldwin et al.\
(1995).  The fit is given in Bottorff et al.\ (1997) and is reproduced here for
convenience.
\begin{equation}
\log{\epsilon(C{~\sc iv})}\propto [\log{\Phi_{18}(H)}]^{0.67}
\end{equation}
Here $\Phi_{18}(H)=\Phi(H)/10^{18}$ where $\Phi(H)$ is the hydrogen ionizing
photon number flux in $\rm{cm^{-2}~s^{-1}}$.  Following Netzer and Laor (1993)
we assume that lines are suppressed with the onset of grain formation at
$\Phi_{18}(H)=1.0$ so we assign zero emissivity to clouds exposed to this flux 
or less.  This is satisfied for
\begin{equation}
8.5\times 10^{-4}R_{G,10}^{2}/L_{45} \approx (v_{G}/\sigma_{G})_{cut}^{4}
\end{equation}
where $R_{G,10}$ is the radius $R_{G}$, 
written in units of 10 light days, $L_{45}$ is the bolometric luminosity in units
of $10^{45}\rm{erg~s^{-1}}$, and $(v_{G}/\sigma_{G})_{cut}$ is the value of
$|v_{G}/\sigma_{G}|$ below which the emissivity is defined to be
zero.  We take $R_{G,10}\approx 1.0$ and $L_{45}\approx 0.27$ (values corresponding to
the Seyfert 1 galaxy NGC~5548, Bottorff et al.\ 1997) so
$(v_{G}/\sigma_{G})_{cut}\approx 0.24$.  For comparison, an example of a quasar is 3C390.3 which has
$R_{G,10}\approx 6.3$ and $L_{45} \approx 1.8$
(Wamsteker et al.\ 1997) giving $(v_{G}/\sigma_{G})_{cut}\approx 0.43$.  The cutoff in 
emissivity is equivalent to truncating the distribution $f(v_{G}/\sigma_{G})$. 
The thick curve in Figure~1 shows $f(v_{G}/\sigma_{G})$ normalized to 
$f(0)=1.0$ and truncated for $|v_{G}/\sigma_{G} |<0.24$.
%
%
\begin{figure}
  \epsfxsize=6.0in\epsfysize=2.0in
  \epsfig{file=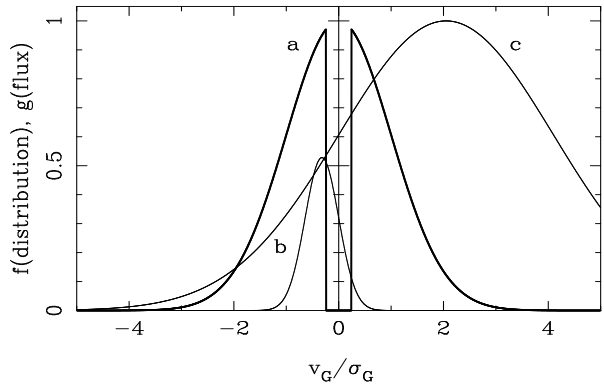, 
          height=60mm, 
          width=80mm,
          bbllx=100pt, bblly=40pt, bburx=650pt, bbury=690pt,
          rheight=52mm, 
          rwidth=80mm, 
          angle=270.0
}
  \caption{Plot of the distribution $f(v_{G}/\sigma_{G})$ normalized to 
$f(0.0)=1.0$ and including a $(v_{G}/\sigma_{G})_{cut}=0.24$ cutoff ((a), thick line).
Also shown is $g(-0.32,v_{G}/\sigma_{G})$ and $g(2.05,v_{G}/\sigma_{G})$ ((b) and (c)
respectively, thin lines).
Both are normalized so that $g(2.05,0.0)=1.0$}
\end{figure}

Our simulation used a total of
84,000 clouds, estimated from typical BLR cloud column densities, particle
densities, and the size and covering fraction of the BLR. 
To simulate MHD wave broadening, each cloud was given a gaussian
line profile centered at a randomly selected value of $v_{G}/\sigma_{G}$ 
denoted as $v_{i}/\sigma_{G}$ and
assigned a dispersion equal in magnitude to $v_{i}/\sigma_{G}$ so as to be
consistent with Equation~7.  The cloud line profile,
$g(v_{i}/\sigma_{G},v_{G}/\sigma_{G})$, is thus
\begin{equation}
g(\frac{v_{i}}{\sigma_{G}},\frac{v_{G}}{\sigma_{G}})\propto\frac{\epsilon(\rm{C{~\sc iv}})}
{\sqrt{4\pi(\frac{v_{i}}{\sigma_{G}})^{2}}}\exp{-\frac{1}{2}\left(\frac{\frac{v_{G}}{\sigma_{G}}-
\frac{v_{i}}{\sigma_{G}}}
{\frac{v_{i}}{\sigma_{G}}}\right)^{2}}
\end{equation}

A cloud with small $|v_{i}/\sigma_{G}|$ (but still larger than
$(v_{G}/\sigma_{G})_{cut}$), has a relatively narrow width and makes a smaller
small flux contribution to the total line profile as compared to clouds with
larger $|v_{i}/\sigma_{G}|$.  For comparison Figure~1 also shows two cloud
profiles, namely $g(-0.32,v_{G}/\sigma_{G})$ and $g(2.05,v_{G}/\sigma_{G})$.
Both have been normalized to $g(2.05,0.0)=1.0$.  It is apparent in the figure
that, individually, clouds with high $|v_{i}/\sigma_{G}|$ outshine those with
low $|v_{i}/\sigma_{G}|$.  For the two cloud profiles shown in Figure~1 the
larger $|v_{i}/\sigma_{G}|$ cloud is 10 times more luminous.  On the other hand
there are many more low velocity clouds than high velocity clouds due to the
distribution $f(v_{G}/\sigma_{G})$.  The resulting profile, due to the
accumulation of all 84,000 clouds is shown in Figure~2 though only 81 percent of
the clouds actually contribute the line profile due to the cutoff.  The Profile
represents the BLR contribution of an AGN emission line.
%
%
\begin{figure}
  \epsfxsize=6.0in\epsfysize=2.0in
  \epsfig{file=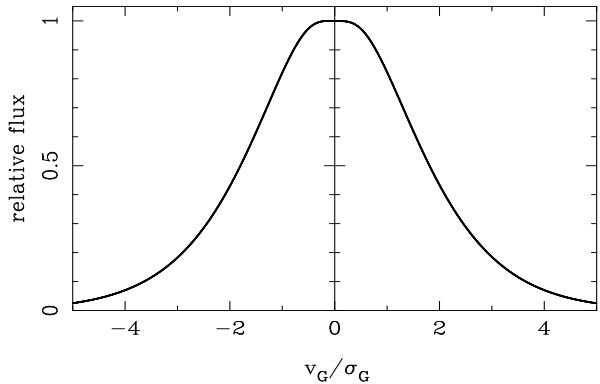, 
          height=60mm, 
          width=80mm,
          bbllx=100pt, bblly=40pt, bburx=650pt, bbury=690pt,
          rheight=52mm, rwidth=80mm, 
          angle=270.0
}
  \caption{Simulated BLR portion of an emission line profile for our simple
  1-D model. For this simulation 84,000 clouds were used with 
  $(v_{G}/\sigma_{G})_{cut} = 0.24$.}
\end{figure}

Analysis of this model shows that lowering the number of clouds to 300 has
little effect on the line profile though it does become somewhat less symmetric
due to the sensitivity of the profile to large individual contributions from
high velocity clouds.  The effect of MHD wave broadening is apparent when the
full width at half maximum (FWHM) of the profile in Figure~2 is compared to the
FWHM of $f(v_{G}/\sigma_{G})$.  The line profile is 1.6 times wider than $f(v_{G}/\sigma_{G})$.
The same model with 84,000 clouds but using $(v_{G}/\sigma_{G})_{cut} = 0.43$ (e.g.,
parameters for 3C390.3) yields a similar profile to the one shown in Figure~2.
It is somewhat wider (the FWHM is 1.7 times larger than the FWHM of $f(v_{G}/\sigma_{G})$)
but has a flat plateau nearly twice as wide.

In the following section we interpret the above results in terms of current
AGN research and make suggestions for future avenues of study.

\subsection[]{Can the Cloud Nature be Determined?}

This simple model brings a whole series of issues to the study of AGN
phenomena.  Attempts to infer cloud numbers
from broad line profiles need to include the possibility of MHD wave
line broadening before making conclusions about whether BLR clouds are
continuous or discrete.  If individual clouds are sought,
the search needs to be redirected away from the line wings and toward the line
core since, by Equation~4, the line broadening, $\sigma_{\rm{B}}$,
will be weaker at smaller values of $v$.   

An alternative approach to AGN cloud counting could be to use a principle
component analysis fitting approach.  The gaussian basis functions used to fit a
profile can be given widths proportional to their offset from zero systemic
velocity.  The minimum number of gaussians required to find an acceptable fit to
the line profiles will be an estimate of the minimum cloud number.  Components
may be tested in AGN that have had extensive spectral monitoring, e.g.
NGC~5548, NGC~4151, 3C390.3 and 3C273.  A sequence of spectra, covering a time
span shorter than the BLR dynamical crossing time but longer than a
characteristic continuum variability timescale, can be fit.  If the component
number and locations in the sequence do not significantly vary, then that would
be evidence in favor of the components being actual clouds.  The ensemble could
be further tested against various kinematic models by tracking detected clouds
over a few crossing times.  There is already evidence to suggest that line
profiles of clouds are a complex amalgam of emitters.  Multiple components are
often required to fit line profiles.  For example, four or five components are
required to fit the H$\beta$ line of Ark~120 (Korista 1992) and H$\beta$ in
NGC~5548 shows three seemingly independent time variable components (Wanders and
Peterson 1996).  We note, for clarity however, that it is not suggested that the
components presented in those papers necessarily represent actual clouds since
the fitting algorithms are designed for efficiency and have no physical meaning.
We do however, wish to emphasize the potential for future reanalysis of
available data.

With regard to magnetic broadening and long term profile variability, consider
the ``shoulders'' of the H$\beta$ line of NGC~5548 reported by Wanders and
Peterson (1996).  They note that shoulders ``do not appear to move
systematically in radial velocity but appear to come and go at approximately
fixed wavelengths.'' In terms of our simple model this behavior can be explained
by the movement of a few relatively high velocity clouds.  Since both the
emission and line width is large in these clouds we would expect that the wings
vary on a time scale of the order of the BLR crossing time, which they estimate
to be $\sim1.6\rm{~yr}$.  This is indeed the case.  The line core, in our
model, however, is dominated by many dimmer clouds.  Thus the profile core will
be relatively stable since the addition or subtraction of a few clouds will not
affect the overall shape of that segment of the line.  In addition, the crossing
time at a lower velocity will be longer than average.  The net result is a
stable line core with wings that occasionally balloon into shoulders.
Shouldering may therefore indicate the movement of a few high velocity extremely
MHD broadened clouds.  An observer might be able to track individual clouds by
subtracting shouldered spectra from a mean spectrum or from a pre or post
shoulder state of a line wing.  Based on our modeling of the line profile for
$\sim 300$ clouds and comparing it with the the amplitude of the shouldering
observed in NGC~5548 there may be only a few hundred clouds in this object
(see also Wanders (1997)).

Current
virial mass estimates based on profile width will be too big. 
Energy equipartition of gravity with the magnetic wave energy results in an 
error of a factor of $\sim1.6^{2}$ $(\sim2.6)$ in the virial mass 
if the FWHM of the line is used.  This is a consequence of Equation~7
($\sigma_{\rm{B}}\approx v_{G}$).  Line profiles are wide because of roughly equal
contributions from the velocity field of the cloud ensemble and the line widths
of individual clouds.  Additional non-kinematic mechanisms used to explain the
extreme wings of broad line profiles, e.g.\ electron scattering (Emmering et
al.\ 1992, Bottorff et al.\ 1997), could also be at work though magnetic
broadening may obviate the need of them to account for the wings of line
profiles.  The widths of clouds in the extreme wings guarantee overlap in
one wing with the profiles of clouds in the other wing and the line core. The
response of the extreme wings to short term changes in the continuum (not to be
confused with the long term kinematic changes discussed above) will be smoothed.
Thus, for short term variability, due to rapid changes in the continuum, our
model predicts a variable central part of the profile and a temporally smoothed response in
the extreme wings.

An observer's orientation with respect to the magnetic field will affect the
observed profile if the field is coherent.  The FWHM of a cloud should be wider in the plane of
oscillation than if viewed along the magnetic field.  In addition, cloud models
need to be modified to include the effects of large non-dissipative
internal motions on radiation transfer within a cloud.  

Finally, dynamical
mechanisms for initiating oscillations need to be investigated.  Spatial wave
amplitudes of
\begin{equation}
A\approx \lambda/2\pi,
\end{equation}
where $\lambda$ is the wavelength, are required to produce transverse
displacement velocities comparable to $\sigma_{\rm{B}}$.
Since the length of a cloud is $l\approx N_{24}/n_{10}\times 10^{14}\rm{cm}$,
where $N_{24}$ is the column density in units of $10^{24}\rm{cm^{-2}}$ and
$n_{10}$ is the cloud particle density in units of $10^{10}\rm{cm^{-3}}$,
the longest wavelength supported in a cloud is of the order $l\approx\lambda$.
Using equation (7) gives the period of oscillation, $T$, is bounded by
\begin{equation}
T\la \frac{\lambda}{\sigma_{\rm{B}}} \approx 3\times 10^{5}\frac{N_{24}}{n_{10}}
\sqrt{\frac{R_{10}}{M_{7}}}\rm{~s}.
\end{equation}
Thus, $T$ is about 4 days \textit{or less} for fiducial values.  We note that many
continuum engines vary on a time scale of this order so radiation pressure could
be important.  The importance of radiation pressure effects in
static magnetic outflows is clearly demonstrated in K\"onigl and Kartje (1994).
Whether or not oscillations can be initiated by it however is still an open
question.

\section[]{Conclusions}

Magnetic confinement and associated MHD wave broadening has 
broad ramifications.
\bigskip

\noindent $\bullet$ Energy equipartition between non-dissipative waves ($1/2\rho
\sigma_{\rm{B}}^{2}$) and the magnetic field ($B^{2}/8\pi$) 
seems to be an inevitable
consequence of magnetically confined gas.  Line broadening is expected everywhere
magnetic fields are found.  In cases where there is also energy equipartition
with gravity, the effects of magnetic line broadening will be extreme.  
\bigskip

\noindent $\bullet$ The spectrum emitted by clouds subject to MHD wave broadening needs to
be reevaluated since the local velocity field will be highly supersonic, not
thermal as previously assumed.  This will have fundamental effects on the
spectrum since line trapping will be far less severe and continuum pumping far
more important (Ferland 1999).  
\bigskip

\noindent $\bullet$ Virial mass estimates will be too large by a factor $\sim 2$ if
the magnetic field is in energy equipartition with gravity.
\bigskip

\noindent $\bullet$ The smallest resolved component in a line may be
considerably broader than its thermal width.  If a component is part of an
ordered flow, then its width can rival that of the flow if the magnetic field is in
energy equipartition with gravity.  This has important implications for current
attempts to detect and count individual cloud elements.  
\bigskip

\noindent $\bullet$ Line profiles broadened by a magnetic field will be more
symmetric than expected by predictions of a kinematic model that uses thermal
broadening only.  Thus the parameters describing the shape of a model line
profile may need to be changed considerably once correct radiative
transfer is applied.

\section*{Acknowledgments}

We thank Jack Baldwin and Kirk Korista for helpful comments.  This work is supported 
through NSF grant AST 96-17083

\end{document}